\documentclass[english,aps,groupedaddress,nofootinbib]{revtex4}
\usepackage[T1]{fontenc}
\usepackage[latin1]{inputenc}
\usepackage{amsmath}
\usepackage{amssymb}
\usepackage{esint}

\makeatletter
\@ifundefined{textcolor}{}
{%
 \definecolor{BLACK}{gray}{0}
 \definecolor{WHITE}{gray}{1}
 \definecolor{RED}{rgb}{1,0,0}
 \definecolor{GREEN}{rgb}{0,1,0}
 \definecolor{BLUE}{rgb}{0,0,1}
 \definecolor{CYAN}{cmyk}{1,0,0,0}
 \definecolor{MAGENTA}{cmyk}{0,1,0,0}
 \definecolor{YELLOW}{cmyk}{0,0,1,0}
 }

\makeatother

\usepackage{babel}

\begin{document}

\title{Condensate Structure of D-particle Induced Flavour Vacuum}

\author{Nick E. Mavromatos}

\affiliation{Department of Physics, King's College London, Strand, London WC2R
2LS, UK}

\author{Sarben Sarkar}

\affiliation{Department of Physics, King's College London, Strand, London WC2R
2LS, UK}

\author{Walter Tarantino}

\affiliation{Department of Physics, King's College London, Strand, London WC2R
2LS, UK}

\date{\today}
\begin{abstract}
It is argued that four Fermi interactions induced by non-perturbative
effects due to scattering of stringy matter from D-particles, D-instantons
and more generally bulk gauge fields in models with large extra dimensions
have in specific situations condensate structure described by flavour
vacua.
\end{abstract}
\maketitle

\section{Introduction}

Recently~\cite{mst2}, within the framework of string-inspired quantum space-time
foam (D-foam)\cite{westmuckett2}, the existence of flavour mixing and induced
oscillations  was deduced based on kinematics of scattering
and potential energy between D-branes in relative motion. This mixing
is over and above the mixing responsible for conventional oscillation
phenomena. It was conjectured in \cite{mst2}, and also in \cite{mst1},\cite{mavrosarkar}, that the resulting state could be described
by a state known as a \emph{flavour vacuum} introduced initially in \cite{Blasone:1995zc}. 

We would like to show here that such a flavour vacuum is a condensate
and can arise explictly through certain four-fermion interactions
that can be induced dynamically by the scattering of strings from
brane defects in the structure of space-time. Depending on the type of string theory considered, these  defects may be either point-like (D0 branes)~\cite{westmuckett2} or compactified Dp-branes, wrapped around p-cycles~\cite{Li:2009tt}. Furthermore, rather
generically, other non-perturbative interactions induced by D-instantons
or Kaluza-Klein modes in brane world scenarios can give rise to similar
four fermion interactions. Hence it is possible that flavour vacua
may well arise from non-perturbative effects which arise in particle
physics beyond the standard model (SM).

\section{D-foam Models and induced Four-Fermion Vector Interactions}

In the D-foam models our Universe is represented
as a Dirichlet three-brane (D3-brane), perhaps after appropriate compactification
of higher-dimensional domain-wall structures of space-time, e.g. as
eight-branes in the model of \cite{westmuckett2}. In such D3 brane
Universe conventional particles propagate as open strings. This 3-brane
propagates in a 10-dimensional bulk space-time containing orientifold
planes, that is punctured by D-particle defects. D-particles cross
the D3-brane world as it moves through the bulk. To an observer on
the D3-brane, these crossings constitute a realization of `space-time
foam' with defects at space-time events due to the D-particles traversing
the D3-brane: we term this structure `D-foam'. When the open strings
encounter D-particles in the foam, their interactions involve energy-momentum
exchange that cause the D-particles to recoil due to scattering

Typical of possibilities beyond SM is the existence of four fermion
interactions that arise in string theory. Tree level four-point scattering
with matter fields living on D-brane interactions lead to four Fermi
interactions\cite{Antoniadis:2000jv}. Such a situation arises in
models where D$3$-branes are inside D$7-$branes \cite{Li:2009tt}.
The D$3$-branes wrap a $3$-cycle thus becoming effectively a D-particle.
The D$7$-branes wrap a $4$-cycle and form the three-space dimensional
world on which the SM particles live. The $3$-cycle is taken to be
$S^{1}\times S^{1}\times S^{1}$. A generic SM particle is represented
by an open string with both ends on the D$7$ brane and is labelled
by Chan-Paton indices $a\bar{b}.$ An open string with one end on
the D$7-$brane and another on the D$3-$brane will be denoted by
ND (denoting Neumann and Dirichlet respectively)~\cite{Antoniadis:2000jv}.
In the D-foam model D-particles are assumed to be uniformly distributed
in the Universe. A D-particle in this foam can capture an $a\bar{b}$
string which then splits into to two $ND$ strings $a\overline{c}$
and $\overline{b}c$. These $ND$ strings scatter off the D-particle
\cite{Li:2009tt} and leave as $\overline{c}a$ and $c\overline{b}$
states. The calculation of these amplitudes was done for massless
fermions~\cite{Li:2009tt} (relevant for massless neutrinos) following
\cite{Antoniadis:2000jv} using string perturbation theory. Such considerations
typically lead to four Fermi interactions of the form \begin{equation}
\mathcal{L}_{{\rm 4-eff}}=\frac{\eta}{M_{s}^{2}}V^{c}\sum_{i,j}\sum_{a,b=L,R}\mathcal{G}_{ab}\overline{\psi_{i}}^{a}\gamma^{\mu}\psi_{i\, a}\,\overline{\psi_{j}}^{\prime\, b}\gamma^{\mu}\psi_{j\, b}^{\prime}\label{fourfermi}\end{equation}
where $V^{c}$ denotes the compactitication volume element to four
space-time dimensions, in units of the string length $\sqrt{\alpha'}=1/M_s$;
the indices $i,j$ refer to fermion species, including flavour, $\mathcal{G}_{ab}$
are numerical coefficients depending on the couplings of the particular
interactions, and are of order $O(1)$ as far as the string scale
$M_{s}$ is concerned, and $L,R$ denote the appropriate chirality
of the spinors $\psi_{i},\psi'_{j}$. The constant $\eta$ depends
on the details of the model considered. (Although the D3/D7 model
allows for four Fermi interactions among charged particles, these are suppressed~\cite{Li:2009tt}
in comparison to neutral fermions.)

The structure of $\mathcal{G}_{ab}$ can be quite involved in more
realistic models since it may depend on $i,j$. Serious attempts have
been made using complex D-brane configurations and orbifold structures
to create low energy effective theory. One such attempt is known as the
intersecting brane models~\cite{Zwiebach}; in this context  and in relation to flavour
changing neutral currents four Fermi interactions have been derived
where $\mathcal{G}_{ab}$ have additional dependence on $i,j$. The effect is
based on a non-perturbative mechanism involving D-instantons. This dependence,
at its root, arises because matter fields can be localised at different
points in the extra dimensions. The calculations are more technically
involved for the case when $a\neq b$ (of special interest to us).
However it was noted that the results can be obtained more easliy
within the brane world field theory where there is typically one brane
and a fifth dimension which is the bulk direction~\cite{Abel:2003fk}
associated with a mass scale much smaller than the Planck scale. The
explicit $i,j$ dependence in such a model is summarised in the appendix
in (\ref{A-coefficient}). $\mathcal{G}$ can be both positive and
negative depending on the details. We will later investigate a simple
model with negative $\mathcal{G}$ for detailed consideration.

We wish to know whether any condensates implied by (\ref{fourfermi})
and its variants are compatible with the flavour vacuum. In order
to answer this we will first investigate the possible condensates
that can arise from Fierz factorisations of the four Fermi interaction. We
will then evaluate such condensate order parameters in the flavour
vacuum. We will then show that a simple four Fermi interaction in
a brane world scenario can be described by such a condensate in the
limit of very small mixing. This should allow the flavour vacuum to
be considered as a possibility chosen by the dynamics and in particular
by dynamics that goes beyond the standard model.

\section{Condensates from Four-Fermi Interactions}

Upon a Fierz rearrangement, the four-fermion interactions (\ref{fourfermi}),
involving both left- and right-handed flavoured fermions, includes
terms : \begin{equation}
\mathcal{L}_{{\rm 4-eff}}\ni\frac{\eta\mathcal{G}V^{c}}{M_{s}^{2}}\sum_{i,j}\left[-(\overline{\psi_{i}}\psi_{j})\,(\overline{\psi_{j}}\psi_{i})+(\overline{\psi_{i}}\gamma_{5}\psi_{j})\,(\overline{\psi_{j}}\gamma_{5}\psi_{i})+\frac{1}{2}(\overline{\psi_{i}}\gamma^{\mu}\psi_{j})\,(\overline{\psi_{j}}\gamma^{\mu}\psi_{i})+\frac{1}{2}(\overline{\psi_{i}}\gamma_{5}\gamma^{\mu}\psi_{j})\,(\overline{\psi_{j}}\gamma_{5}\gamma^{\mu}\psi_{i})\right]~,\label{fierz}\end{equation}
 where for brevity and concreteness we have assumed one type of Non-Standard-Model
interaction among neutrinos due to the intermediate string states,
 and that the spinors have both left and right-handed
components, \emph{i.e.} $\psi_{i}=\begin{pmatrix}\psi_{L}\\
\psi_{R}\end{pmatrix}_{i}~,\quad i=1,\dots N$ where $N$ is the number of flavours. The reader should notice that
in the Fierz transformation (\ref{fierz}) the scalar and pseudoscalar
terms have the opposite sign ( but the same sign for the vector and
axial vector terms). This implies that for fields of the same chirality
the scalar and pseudo scalar contributions would cancel. Hence, if
this induced four -Fermi interaction was to contribute, it is necessary
to have both right and left-handed fermions. In the case of neutrinos
this would imply the necessity of right-handed sterile neutrinos.

This has important consequences for the possibility of formation of
fermion scalar condensates $\langle\overline{\psi}_{i}\psi_{j}\rangle$.
Provided $\mathcal{G}$ is negative, from (\ref{fierz}), we would
obtain \emph{attractive} interactions in the scalar channel of the
form \begin{equation}
\mathcal{L}_{{\rm 4-eff}}\ni-\frac{\eta\mathcal{G}V^{c}}{M_{s}^{2}}\sum_{i,j}\langle(\overline{\psi}_{i}\psi_{j})\rangle(\overline{\psi}^{j}\psi^{i})~.\label{fierz2}\end{equation}
 The condensates would lead to dynamical mass matrix contributions:
\begin{equation}
m_{ij}=\frac{\eta\mathcal{G}V^{c}}{M_{s}^{2}}\langle\overline{\psi}_{i}\psi_{j}\rangle,\quad i,j=1,\dots N~.\label{massmatrixdfoam}\end{equation}
For models with \emph{extra dimensions} and flavours, located at different
positions in the bulk, it is easy to calculate $\mathcal{G}$~\cite{Abel:2003fk} and
find that it can be negative. This is given in an appendix (\emph{cf.} (\ref{A-coefficient})). In principle
the other terms in (\ref{fierz}) can contribute to a condensate.
The nature of the condensates supported by the flavour vacuum will
be investigated next.

\section{Condensates for the Flavour Vacuum }

The well-known Pontecorvo mixing transformation~\cite{Bilenky} at
the level of fields (rather than states) has the form\begin{equation}
\psi_{\alpha}\left(x\right)=T\left(\theta\right)_{\alpha j}\psi_{j}\label{Pontecorvo}\end{equation}
 where $T\left(\theta\right)_{\alpha j}$ is a c\emph{-number} (Pontecorvo
flavour rotation matrix). For two flavour mixing $T$ is a $2\times2$
rotation through an angle $\theta$ (i.e. $SO\left(2\right))$ matrix.
This is the customary way of dealing with flavour in the SM. The work
of Blasone and Vitiello (BV)~\cite{Blasone:1995zc} made the ingenious suggestion
that (\ref{Pontecorvo}) is replaced by  implementation of mixing
via a quantum operator $G_{\theta}$ rather than a c-number matrix,
i.e. \begin{equation}
\psi_{a,b}\left(x\right)=G_{\theta}^{-1}\psi_{1,2}\left(x\right)G_{\theta}.\label{BVtransform}\end{equation}
The operator $G_{\theta}$ is however formal and takes states from
the massive Fock space to another Fock space which in the thermodynamic
limit is orthogonal~\cite{Blasone:1995zc} to the massive Fock space.
For massive Majorana (Dirac) fields $\psi_{i},$with masses $m_{i}$($i=1,2$),
$G_{\theta}$, for example, is given by \begin{equation}
G_{\theta}\left(t\right)=\exp\left(\theta\int d^{3}x\left[\psi_{1}^{\dagger}\left(\overrightarrow{x},t\right)\psi_{2}\left(\overrightarrow{x},t\right)-\psi_{2}^{\dagger}\left(\overrightarrow{x},t\right)\psi_{1}\left(\overrightarrow{x},t\right)\right]\right).\label{BVoperator}\end{equation}
We will show that it is dynamics that chooses between the inequivalent
representations in (\ref{Pontecorvo}) and (\ref{BVoperator}). As
we will show the latter implies a condensate vacuum state whereas
the former does not. In terms of the massive Fock vacuum $\left|0\right\rangle _{1,2}$,
the flavour vacuum $\left|0\right\rangle _{f}$ is given by \begin{equation}
\left|0\right\rangle _{f}\equiv G_{\theta}^{-1}\left|0\right\rangle _{1,2}\label{FlavourVacuum}\end{equation}
The condensate nature of the flavour vacuum can be probed by calculating
the expectation values of the various operators that appear in the
Fierz decomposition (\ref{fierz}). They can be calculated straightforwardly
by writing the usual plane-wave expansion of the massive Majorana
fermionic fields~\cite{Itzykson:1980rh} \begin{equation}
\psi_{i}\left(x\right)=\sum_{r=1,2}\int\frac{d^{3}k}{\left(2\pi\right)^{3/2}}\left(a_{i}^{r}\left(\overrightarrow{k}\right)u_{i}^{r}\left(\overrightarrow{k}\right)e^{-i\omega_{i}\left(k\right)t}+v_{i}^{r}\left(-\overrightarrow{k}\right)a_{i}^{r\dagger}\left(-\overrightarrow{k}\right)e^{i\omega_{i}\left(k\right)t}\right)e^{i\overrightarrow{k}.\overrightarrow{x}}\label{FreeField}\end{equation}
with $\omega_{i}\left(k\right)=\sqrt{k^{2}+m_{i}^{2}}$ and $k^{2}=\overrightarrow{k}$$^{2}$.
The spinors $u_{i}^{r}\left(\overrightarrow{k}\right)$and $v_{i}^{r}\left(\overrightarrow{k}\right)$are
determined by \[
\begin{cases}
\begin{array}{c}
\left(\gamma^{0}\omega_{i}\left(k\right)+\overrightarrow{\gamma}.\overrightarrow{k}-m_{i}\right)u_{i}^{r}\left(\overrightarrow{k}\right)=0\\
u_{i}^{r\dagger}\left(\overrightarrow{k}\right)u_{i}^{s}\left(\overrightarrow{k}\right)=\delta^{rs}\\
v_{i}^{r}\left(\overrightarrow{k}\right)=\gamma^{0}C\, u_{i}^{r}\left(\overrightarrow{k}\right)^{*}\end{array}\end{cases}\]
and the charge conjugation operator $C=-i\sigma^{1}\otimes\sigma^{2}$ (where
$\sigma^{i}$are the Pauli spin matrices). The flavour fields in (\ref{BVtransform})
can be expanded as in (\ref{FreeField}) but with $a_{i}^{r}$ replaced
by $a_{\alpha}^{r}$with $\alpha=a,b$. On using (\ref{BVtransform})
and (\ref{BVoperator}) it is then readily shown that we have the
Bogolubov relations\begin{equation}
a_{a}^{r}\left(\overrightarrow{k},t\right)=\cos\theta a_{1}^{r}\left(\overrightarrow{k}\right)-\sin\theta\sum_{s}\left(W^{rs}\left(\overrightarrow{k},t\right)a_{2}^{s}\left(\overrightarrow{k}\right)+Y^{rs}\left(\overrightarrow{k},t\right)a_{2}^{s\dagger}\left(-\overrightarrow{k}\right)\right)\label{Bogolubov1}\end{equation}
\begin{equation}
a_{b}^{r}\left(\overrightarrow{k},t\right)=\cos\theta a_{2}^{r}\left(\overrightarrow{k}\right)+\sin\theta\sum_{s}\left(W^{sr*}\left(\overrightarrow{k},t\right)a_{1}^{s}\left(\overrightarrow{k}\right)+Y^{sr}\left(-\overrightarrow{k},t\right)a_{1}^{s\dagger}\left(-\overrightarrow{k}\right)\right)\label{Bogolubov2}\end{equation}
where \[
W^{rs}\left(\overrightarrow{k},t\right)\equiv\frac{1}{2}\left(u_{2}^{r\dagger}\left(\overrightarrow{k}\right)u_{1}^{s}\left(\overrightarrow{k}\right)++v_{2}^{s\dagger}\left(\overrightarrow{k}\right)v_{1}^{r}\left(\overrightarrow{k}\right)\right)e^{i\left(\omega_{1}-\omega_{2}\right)t}\]
and\[
Y^{rs}\left(\overrightarrow{k},t\right)\equiv\frac{1}{2}\left(u_{2}^{r\dagger}\left(\overrightarrow{k}\right)v_{1}^{s}\left(-\overrightarrow{k}\right)+u_{1}^{s\dagger}\left(-\overrightarrow{k}\right)v_{2}^{r}\left(\overrightarrow{k}\right)\right)e^{i\left(\omega_{1}+\omega_{2}\right)t}.\]
These relations allow the computation of condensates. For operators
\emph{normal ordered} with respect to the massive vacuum we find:
\begin{equation}\label{ijcond}
_{f}\left\langle 0\right|\overline{\psi}_{i}\psi_{j}\left|0\right\rangle _{f}=\frac{1}{2}\sin2\theta\,\int^{\Lambda}dk\frac{k^{2}}{\pi^{2}}\left(\frac{m_{j}}{\omega_{j}}-\frac{m_{i}}{\omega_{i}}\right)\qquad\mathrm{for}\, i\neq j.
\end{equation}
As in treatments of condensates there needs to be input from a more
fundamental theory to determine the cut-off $\Lambda$, necessary
to make the integral finite. When $i=j$ the condensate expectation
vanishes. Some arguments in favour of a cut-off 
\begin{equation}\label{cutoff}
\Lambda = \overline{m} +\mathcal{O}((\delta m)^2/\overline{m})~, 
\quad \overline{m}  \equiv \frac{m_1 + m_2}{2}, \quad \delta m \equiv m_1 - m_2~,
\end{equation}
for two-flavoured systems, that we consider here for simplicity, 
have been given in \cite{mst1},\cite{mavrosarkar},\cite{Barenboim:2004ev} based on particle production characteristics of the flavour vacuum in the context of space-time foam expanding Universes. From (\ref{ijcond}), (\ref{cutoff}), and small $\delta m \ll \overline{m}$, we then obtain:
\begin{equation}\label{ijcond2}
_{f}\left\langle 0\right|\overline{\psi}_{i}\psi_{j}\left|0\right\rangle _{f} \simeq 
2.5 \times 10^{-3} \, {\rm sin}(2\theta) \, \overline{m} \, (m_1^2 - m_2^2) + \dots ~,
\end{equation}
where the $\dots$ indicate higher orders in $\delta m$. 

The other possible
condensates implied by (\ref{fierz}) vanish (\emph{e.g.} $_{f}\left\langle 0\right|\overline{\psi_{i}}\gamma^{\mu}\psi_{j}\left|0\right\rangle _{f}=0)$.
Hence the flavour vacuum is a pure mixing condensate. We shall show
that in the next section that a simple brane world field theory reproduces
the same structure. It should be stressed here that a D-particle foam
may have other effects \cite{mst1},\cite{mavrosarkar} beyond those
that it shares with other non-perturbative effects such as D-instantons
and brane world field theories. This can be seen from the string amplitude
considered at the beginning in our D3/D7 discussion. The capture process
of the string with both ends on the D7 brane by the D-particle was
excluded. This exclusion also ignores interesting effects of particle
production and localy non-flat backgrounds.We will come back to this
issue in a future publication.

\section{Mixing Condensate in Brane Field Theory Model} 

We shall consider a simple (toy) field theoretic four dimensional brane
model with a large extra dimension denoted by $z$ where the generation of a mixing condensate due to four fermion interaction can be demonstrated . The brane will
be taken to be at $z=0$.The model has two species of fermions located
differently with respect to the extra dimension of size $\mathcal{R}$.
One species, $\psi$ is on the brane while the other $\Psi$ is off
the brane. There is no preferred place for us to locate this second
fermion and so it will be allowed to explore the bulk dimension~\cite{Abe:2000ny}.
The bulk fermions interact through the exchange of gravitons and gauge
bosons ( as well as associated Kaluza-Klein modes) and so there are
induced four fermion interactions. This model has the virtue of displaying
a mixing condensate dynamically and, perhaps even more interestingly,
the possibility that the mass scale of the condensate order parameter
is much smaller than the inherent mass scale $\frac{1}{\mathcal{R}}$ (the order of a typical Kaluza-Klein mass). The non-zero condensate starts to appear at a second
order phase transition at $\mathcal{R}=\mathcal{R}_{c}$. For $\mathcal{R}<\mathcal{R}_{c}$ a
non-zero condensate is not generated dynamically while for $\mathcal{R}$
just above $\mathcal{R}_{c}$the condensate mass scale is very small.
In order to have a controlled method of calculating the fermion condensate,
the fields $\Psi$and $\psi$ come in $N_{f}$ different fermion species
allowing a calculation to leading order in $\frac{1}{N_{f}}.$ It
is assumed that the fields $\Psi$ and $\psi$ have the same charge
$g$ with respect to a $U\left(1\right)$gauge field. Just as discussed
in the Appendix, on integrating over the heavy Kaluza-Klein modes,
the following simplified lagrangian\cite{Abe:2000ny} can be obtained
\[
L=\overline{\Psi}i\gamma^{M}\partial_{M}\Psi+\left[\overline{\psi}i\gamma^{\mu}\partial_{\mu}\psi-\overline{G}\left(\overline{\Psi}\gamma_{\mu}\Psi\right)\left(\overline{\psi}\gamma^{\mu}\psi\right)\right]\delta\left(z\right)\]
 where $\overline{G}\propto\mathcal{R}g^{2}$ and is $O\left(\frac{1}{M_{n}^{2}}\right)$.We
will use our earlier Fierz transformation and make the chiral rotation
$\Psi\rightarrow\exp\left(i\frac{\pi}{4}\gamma^{5}\right)\Psi$ and
$\psi\rightarrow\exp\left(i\frac{\pi}{4}\gamma^{5}\right)\psi$. Scalar
and pseudoscalar composites $\sigma=\overline{\Psi}\psi$ and $\varpi=\overline{\Psi}i\gamma_{5}\psi$are
introduced. The usual Kaluza-Klein expansion of the bulk fermion can
be made, i.e. \[
\Psi\left(x,z\right)=\frac{1}{\sqrt{2\pi R}}\sum_{n=-\infty}^{\infty}\Psi_{n}\left(x\right)\chi_{n}\left(z\right)\]
 where \[
\chi_{n}\left(z\right)=\left\{ \begin{array}{cc}
1, & \mbox{when \ensuremath{n=0}}\\
\sqrt{2}\cos\left(\frac{nz}{R}\right), & \mbox{when \ensuremath{n=1,2,\ldots}}\\
\sqrt{2}\sin\left(\frac{nz}{R}\right), & \mbox{when \ensuremath{n=-1,-2,\ldots}}\end{array}\right.\]
 The Lagrangian then reduces to \[
L=\overline{\Xi}\left(M+i\partial^{\mu}\gamma_{\mu}\right)\Xi\]
 where \[
\Xi^{t}=\left(\psi,\Psi_{0},\Psi_{1},\Psi_{-1},\Psi_{2},\Psi_{-2},\ldots\right)\]
 and \[
M=\left(\begin{array}{ccccccc}
0 & m^{*} & m^{*} & m^{*} & m^{*} & m^{*} & \cdots\\
m & 0 & 0 & 0 & 0 & 0 & \cdots\\
m & 0 & \frac{1}{R} & 0 & 0 & 0 & \cdots\\
m & 0 & 0 & -\frac{1}{R} & 0 & 0 & \cdots\\
m & 0 & 0 & 0 & \frac{2}{R} & 0 & \cdots\\
m & 0 & 0 & 0 & 0 & -\frac{2}{R} & \cdots\\
\vdots & \vdots & \vdots & \vdots & \vdots & \vdots & \ddots\end{array}\right)\]
 with $m=\sqrt{\overline{G}/2\pi R}\sigma.$ In a $1/N_{f}$ expansion
the effective potential for $\sigma$ ( with the customary cut-off
$\Lambda$) is \[
V\left(\sigma\right)=\left|\sigma\right|^{2}-\frac{1}{2\pi^{2}}\int_{0}^{\Lambda}dx\, x^{3}\ln\left[x^{2}+\left|m\right|^{2}\left(\pi xR\right)\coth\left(\pi xR\right)\right]\]
 \[
-\frac{1}{2\pi^{2}}\sum_{j=1}^{\infty}\int_{0}^{\Lambda}dxx^{3}\left[x^{2}+\left(\frac{j}{R}\right)^{2}\right].\]
 The minimum of $V\left(\sigma\right)$ is given by the solution of
\[
1-\frac{g^{2}}{2\pi^{2}}\int_{0}^{\Lambda}dx\,\frac{x^{3}}{2x\tanh\left(\pi xR\right)+g^{2}\left|\left\langle \sigma\right\rangle ^{2}\right|}=0.\]
 A non-zero solution for $\sigma$ has been found, numerically, for
suitable $g$ and $R$~\cite{Chang:1999nh}. Furthermore, $\sigma$
is an order parameter for a second order phase transition (approximate
analytic solutions for small $\sigma$ can be found by deriving a
first order differential equation for $\sigma\left(\Lambda\right)$.)
Hence, in this case we have explicitly demonstrated a non-zero off-diagonal
element of the two-fermion mass matrix. The analysis can be repeated
for gravitational Kaluza-Klein modes~\cite{Han:1998sg} with an equivalent
form for the effective Lagrangian and similar conclusion.

\section{Conclusions}

We have shown that from interactions beyond the standard model fermionic
condensates can be formed which may be described by flavour vacua.
The existence of the condensates is very much dependent on the details
of the model. Hence the flavour vacuum is not a procedure that is
a universal panacea when mixing phenomena are involved. The nature
of the dynamical interactions responsible for the mixing will be important
for describing whether such a discussion is possible. The discussion
that we have given in this paper indicates that it may be a way to
describe mixing when we have sufficiently large extra dimensions with
long range fields in the bulk.

\acknowledgments WT thanks King's College London for a postgraduate
studentship when the majority of this work was performed.

\appendix
%dummy comment inserted by tex2lyx to ensure that this paragraph is not empty
\setcounter{equation}{0} \global\long\def\theequation{A\arabic{equation}}

\section{Brane World Calculation of $\mathcal{G}$}

\label{a1}

The generic four-Fermi interactions obtained from a string instanton
analysis is readily obtained in a field theoretic brane world setting~\cite{Abel:2003fk}.
We will consider for simplicity a five-dimensional model with an Abelian
gauge field $A_{M}\,(M=0,\ldots,4)$ interacting with four dimensional
fields $\psi_{j}$located at $z=z_{j}$ ($z$ being the co-ordinate
in the bulk) and $j$ is a flavour index (with now a geometrical interpretation).
The dimension represented by $z$ is compact and of length $L$. This
situation can arise from a $\mbox{D\ensuremath{4}}-$brane which wraps
around a circle of diameter $\frac{L}{\pi}$. The lagrangian $\mathcal{L}$is
given as \[
\mathcal{L}=-\frac{1}{4}F^{MN}F_{MN}+i\overline{\psi}_{j}\gamma^{\mu}D_{\mu}\psi_{j}\delta\left(z-z_{j}\right)\]
 where $F_{MN}=\partial_{M}A_{N}-\partial_{M}A_{N}$ , $D_{M}=\partial_{M}+ig\sqrt{L}A_{M}$
and $\mu,\nu=0,\ldots,3$. (The abelian nature of the field is not
essential and has been chosen for simplicity.) Now considering a Fourier
series in $z$ we have\begin{equation}
A_{\mu}\left(x,z\right)=\frac{1}{\sqrt{L}}A_{\mu}^{\left(0\right)}\left(x\right)+\sqrt{\frac{2}{L}}\sum_{n=1}^{\infty}\left(\cos\frac{2\pi nz}{L}A_{\mu}^{\left(n\right)}\left(x\right)+\sin\frac{2\pi nz}{L}\tilde{A}_{\mu}^{\left(n\right)}\left(x\right)\right).\label{KaluzaKlein}\end{equation}
 We have ignored $A_{4}$ since it does not couple to the fermions
which we have confined to the brane. The fields $A_{\mu}^{\left(n\right)}\mathrm{and}$$\tilde{A}_{\mu}^{\left(n\right)},\, n\geq1,$
are the $n$-th level massive KK fields with mass $M_{n}\left(=\frac{2\pi n}{L}\right)$.
On integrating over $z$ in the action we have an effective Lagrangian
$\mathcal{L}$ given by \[
\mathcal{L}=-\frac{1}{4}\left[F_{\mu\nu}^{\left(0\right)2}+\sum_{n=1}^{\infty}\left(F_{\mu\nu}^{\left(n\right)2}+F_{\mu\nu}^{'\left(n\right)2}\right)\right]+\frac{1}{2}\sum_{n=1}^{\infty}M_{n}^{2}\left(A_{\mu}^{\left(n\right)2}+\tilde{A}_{\mu}^{\left(n\right)2}\right)\]
 \begin{equation}
+i\overline{\psi}_{j}\gamma^{\mu}\left[\partial_{\mu}+igA_{\mu}^{\left(0\right)}+ig\sqrt{2}\sum_{n=1}^{\infty}\left(A_{\mu}^{\left(n\right)}\cos\frac{2\pi nz_{j}}{L}+\tilde{A}_{\mu}^{\left(n\right)}\sin\frac{2\pi nz_{j}}{L}\right)\right]\psi_{j}\label{EffLag}\end{equation}
 where the argument $x$ has been suppressed and the effective $4\mbox{-dimensional coupling}$
$g$ is given in terms of the five dimensional coupling $g_{5}$by
$g=g_{5}/\sqrt{L}$. There is a flavour dependence in the interactions
through $z_{j}$ (i.e. the couplings with $A_{\mu}^{\left(n\right)}$
and $\tilde{A}_{\mu}^{\left(n\right)}$ are $\sqrt{2}g\cos\left(M_{n}z_{j}\right)$and
$\sqrt{2}g\sin\left(M_{n}z_{j}\right)$ respectively); the KK contribution
to the interaction has thus the form \[
\mathcal{L}_{n}=A_{\mu}^{\left(n\right)}J^{\left(n\right)\mu}+\tilde{A}_{\mu}^{\left(n\right)}\tilde{J}^{\left(n\right)\mu}\]
 where $J^{\left(n\right)\mu}=\sqrt{2}g\sum_{j}\cos\left(M_{n}z_{j}\right)\overline{\psi}_{j}\gamma^{\mu}\psi_{j}$
and $\tilde{J}^{\left(n\right)\mu}=\sqrt{2}g\sum_{j}\sin\left(M_{n}z_{j}\right)\overline{\psi}_{j}\gamma^{\mu}\psi_{j}$.
Integration over the $A_{\mu}^{\left(n\right)}$and $\tilde{A}_{\mu}^{\left(n\right)}$
leads to a current-current (\emph{i.e}. a four-Fermi) interaction
(which is not diagonal in flavour) \[
-\frac{1}{2}\sum_{n}\frac{J^{\left(n\right)\mu}J_{\mu}^{\left(n\right)}+\tilde{J}^{\left(n\right)\mu}\tilde{J}_{\mu}^{\left(n\right)}}{M_{n}^{2}}\]
 where $M_{n}=\frac{2\pi n}{L}$. The amplitudes have the form \[
c_{ik}\left(\bar{\psi}_{iL}\gamma^{\mu}\psi_{iL}\right)\left(\bar{\psi}_{kL}\gamma_{\mu}\psi_{kL}\right)\]
 where \begin{equation}
c_{ik}=2g^{2}\sum_{n=1}^{\infty}\frac{\cos\left(M_{n}\left(z_{i}-z_{k}\right)\right)}{M_{n}^{2}}\label{A-coefficient}\end{equation}

\[
=\left(\frac{gL}{2\pi}\right)^{2}\left[\mathrm{Li_{2}\left(\exp\left(\mathit{\frac{2\pi i\left(z_{i}-z_{k}\right)}{L}}\right)\right)+Li_{2}\left(\exp\left(\mathit{\frac{-2\pi i\left(z_{i}-z_{k}\right)}{L}}\right)\right)}\right]\]
 and $\mathrm{Li_{n}}\left(z\right)\mathit{=}\sum_{k=1}^{\infty}\frac{z^{k}}{k^{n}}.$
If a higher dimensional bulk space had been adopted then instead of
$n$ we would require a vector of integers $\overrightarrow{n}$.

\label{a2}

\appendix
%dummy comment inserted by tex2lyx to ensure that this paragraph is not empty
\setcounter{equation}{0} \global\long\def\theequation{A\arabic{equation}}

\end{document}